\documentclass[pdftex,twocolumn,epjc3]{svjour3}

\RequirePackage[T1]{fontenc}

\smartqed

\RequirePackage{graphicx}
\RequirePackage{mathptmx} 
\RequirePackage{flushend}
\RequirePackage[numbers,sort&compress]{natbib}
\RequirePackage[colorlinks,citecolor=blue,urlcolor=blue,linkcolor=blue]{hyperref}

\usepackage{amsmath} 
\usepackage{dsfont} 

\begin{document}

\title{Methods to Select Soft Diffraction Dissociation at the LHC}

\author{Emily Nurse\thanksref{addr1}
        \and
        Sercan Sen\thanksref{e1,addr2,addr3} 
}

\thankstext{e1}{e-mail: Sercan.Sen@cern.ch}

\institute{Department of Physics and Astronomy, University College London, London, UK \label{addr1}
          \and
          Department of Physics Engineering, Hacettepe University, Ankara, Turkey\label{addr2}
          \and
		  Department of Physics and Astronomy, University of Iowa, Iowa City, USA\label{addr3}}

\maketitle
\begin{abstract}
Diffractive events at hadron colliders are typically characterised by a region of the detector
without particles, known as a rapidity gap. In order to observe diffractive events in this way,
we consider the pseudorapidity acceptance in the forward region of the ATLAS and CMS detectors
at the Large Hadron Collider (LHC) and discuss the methods to select soft diffractive dissociation
for $pp$ collisions at $\sqrt{s}=$ 7 TeV. We showed that in the limited detector rapidity acceptance, 
it is possible to select diffractive dissociated events by requiring a rapidity gap in the event, 
however, it is not possible to distinguish single diffractive dissociated events from double diffractive 
dissociated events with a low diffractive mass.
\end{abstract}

\section{Introduction}

The soft diffractive processes at the Large Hadron Collider, LHC, are important for understanding
non-perturbative QCD effects and they also constitute a significant fraction of the total proton-proton 
($pp$) cross-section~\cite{Albrow:2006xt, Albrow:2008az}. Therefore, the measurement of the main characteristics of 
diffractive interactions are essential to improve our understanding of $pp$ collisions. However, the 
modelling of diffraction is still mainly generator dependent and there is no unique, agreed upon, experimental 
definition of diffraction~\cite{Khoze:2010by, Ryskin:2007qx, Gotsman:1993ux, Abe:1993wu, Affolder:2001vx, CMS:2011ysa, 
ATLAS:2010kza, Aad:2012pw, Abelev:2012sea}.

While the physics of diffractive dissociation at the LHC is very important, the detector capabilities 
in the forward region are limited. In this paper, we study the methods to select soft diffraction 
dissociation by considering the forward rapidity coverage of the LHC experiments ATLAS and CMS~\cite{Aad:2008zzm, Chatrchyan:2008aa}.

\section{Event Classification}

In proton-proton (or more generally hadron-hadron) scattering, interactions can be classified 
as either elastic or inelastic by the characteristic signatures of the final states.
Furthermore, it is conventional to divide inelastic processes into diffractive and
non-diffractive parts. In the theoretical concept, hadronic diffractive dissociation is principally 
explained to be mediated by the exchange of the Pomeron, which carries the quantum numbers of the vacuum;
thus, the initial and final states in the scattering process have the same quantum numbers.
If the Pomeron exchange process is additionally associated with a hard scattering (such as the production 
of jets, b-quark, W boson etc), the process known as hard diffractive, otherwise it's soft diffractive dissociation.
Introductory reviews on the area can be found in Refs. \cite{Donnachie:2002en, Barone:2002cv, d'Enterria:2008is}.
Diffractive events at hadron colliders can be classified into the following categories; single, 
double diffractive dissociation and central diffraction (a.k.a. ``Double Pomeron Exchange''), with higher 
order ``multi Pomeron'' processes \cite{Engel:1995sb, Ryskin:2009qf}. Thus, the total proton-proton cross-section can 
be written as the following series
\begin{equation}
      \begin{split}
        \sigma_{total} &= \sigma_{elastic} + \sigma_{inelastic} \\
                   &= \sigma_{elastic} + \sigma_{ND} + \sigma_{diffractive} \\
                   &= \sigma_{elastic} + \sigma_{ND} + \sigma_{SDD} + \sigma_{DDD} + \sigma_{DPE} + \sigma_{MPE} \\
      \end{split}
\end{equation}
where ND is Non-Diffractive processes, SDD (DDD) is Single (Double) Diffraction Dissociation, DPE corresponds
to the Double Pomeron Exchange and MPE refers to the Multi Pomeron Exchange.
Diffractive processes together with the elastic scattering represent about 50\% of the total $pp$ 
cross-section~\cite{Abelev:2012sea, Antchev:2013gaa, Oljemark:2013wsa}.

\begin{figure}[!ht]
  \begin{center}
    \includegraphics[width=0.45\textwidth]{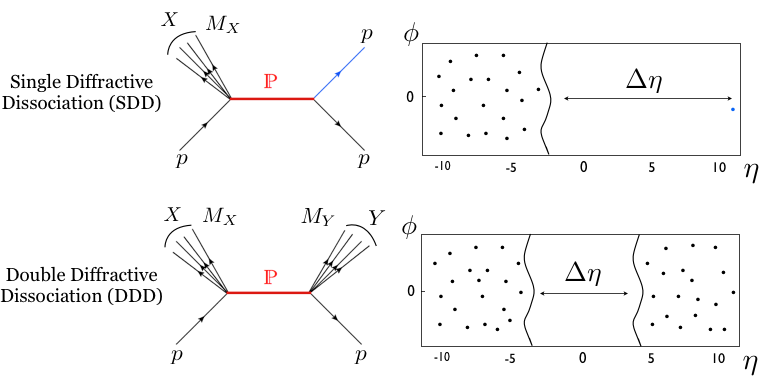}\\
    \caption{Illustration of a single (top) and double (bottom) diffractive dissociative event
             in which a Pomeron ($\mathds{P}$) is exchanged in a $pp$ collision. $M_{X}$ and $M_{Y}$ 
			 are the invariant masses of the dissociated systems $X$ and $Y$, respectively. In single 
			 diffractive dissociation $M_{Y}=m_{p}$ where $m_{p}$ is the mass of the intact proton. 
			 $\Delta\eta$ refers to the size of the large rapidity gap.}
\label{fig:Events}
  \end{center}
\end{figure}

In single diffractive dissociation one of the incoming protons dissociates into a ``low mass'' system (a system 
of particles with low invariant mass w.r.t. the centre of mass energy of the collision) while in double 
diffractive dissociation both of the incoming protons dissociates into ``low mass'' systems as 
represented in Fig.~\ref{fig:Events}.

Diffractive events are classified by a large gap in the pseudorapidity\footnote{The 
pseudorapidity of a particle is defined as $\eta$ = ln tan$(\theta/2)$ where $\theta$ is the polar
angle w.r.t. the beam direction ($z$ axis) and rapidity is $y$ = ln [$(E+p_{z})/(E-p_{z})$] where $p_{z}$ is the
longitudinal momentum of the particle. Pseudorapidity and rapidity are equal for massless particles.}
distribution of final state particles. The large rapidity gap can be defined as
the difference between the rapidity of the diffractively scattered proton and that
of the particle closest to it in (pseudo)rapidity. However, the existing ATLAS and CMS
detectors are not well suited for measuring the forward rapidity gaps.
Therefore, from the experimental point of view, rapidity gaps should be defined by a total absence of particles
in a particular interval of pseudo-rapidity. The large rapidity gap, $\Delta\eta$, is the largest rapidity 
gap between those rapidity gaps in a final state and determines the type of the diffraction process.

\section{Diffractive Kinematical Variables}

\subsection{Fractional Longitudinal Momentum Loss}

In single diffractive collisions, one of the two incident protons emits a Pomeron and remains intact
by loosing a few percent of its initial longitudinal momentum. The fractional longitudinal momentum 
loss of the intact proton is related to the momentum fraction taken by the Pomeron,
    \begin{equation}
        \xi_{X}=1-\left(\frac{{p_{z}^\mathrm{final}}}{{p_{z}^\mathrm{initial}}}\right)
    \end{equation}
where ${p_{z}^\mathrm{final}}$ is the final and ${p_{z}^\mathrm{initial}}$ is the initial longitudinal momentum of the proton.
The Pomeron scatters with the other beam proton and the proton dissociates into a system of particles
with low invariant mass, $M_{X}$. DDD processes are described by the invariant masses $M_{X}$ and $M_{Y}$ of 
the dissociation systems $X$ and $Y$, respectively, as shown in Fig.~\ref{fig:Events}.
The fractional longitudinal momentum loss of the proton can be determined by measuring the invariant mass of the 
dissociated system(s) given as
\begin{equation}
        \xi_{X} = \left(\frac{M_{X}}{\sqrt{s}}\right)^2 \,\,\,\,\,\,\,\,\,,\,\,\,\,\,\,\,\,\, \xi_{Y} = \left(\frac{M_{Y}}{\sqrt{s}}\right)^2
\end{equation}
where $\sqrt{s}$ is the center-of-mass energy for proton-proton collisions.
In the following, the convention $M_{X}>M_{Y}$ is adopted and $\xi_{X}$ is referred to as $\xi$.

There are several other approaches to determine $\xi$ experimentally.
One of the methods is to detect the scattered proton by using e.g Roman pots~\cite{Anelli:2008zza, Schoeffel:2009aa} and
to measure the final momentum of the proton.
The fractional longitudinal momentum loss $\xi$ can also be reconstructed by using the final state particles as 
shown in the following equation.
\begin{equation}
    \xi=\sum_{i}\frac{1}{\sqrt{s}}{E_{T_{i}}e^{\eta_{i}}}
\end{equation}
where ${E_{T_{i}}}$ is the transverse energy and ${\eta_{i}}$ is the pseudorapidity of the $ith$ particle.
We do not consider this way of calculating $\xi$ in this paper.

\subsection{Diffractive Mass}

\begin{figure}[!ht]
  \begin{center}
    \includegraphics[width=0.42\textwidth]{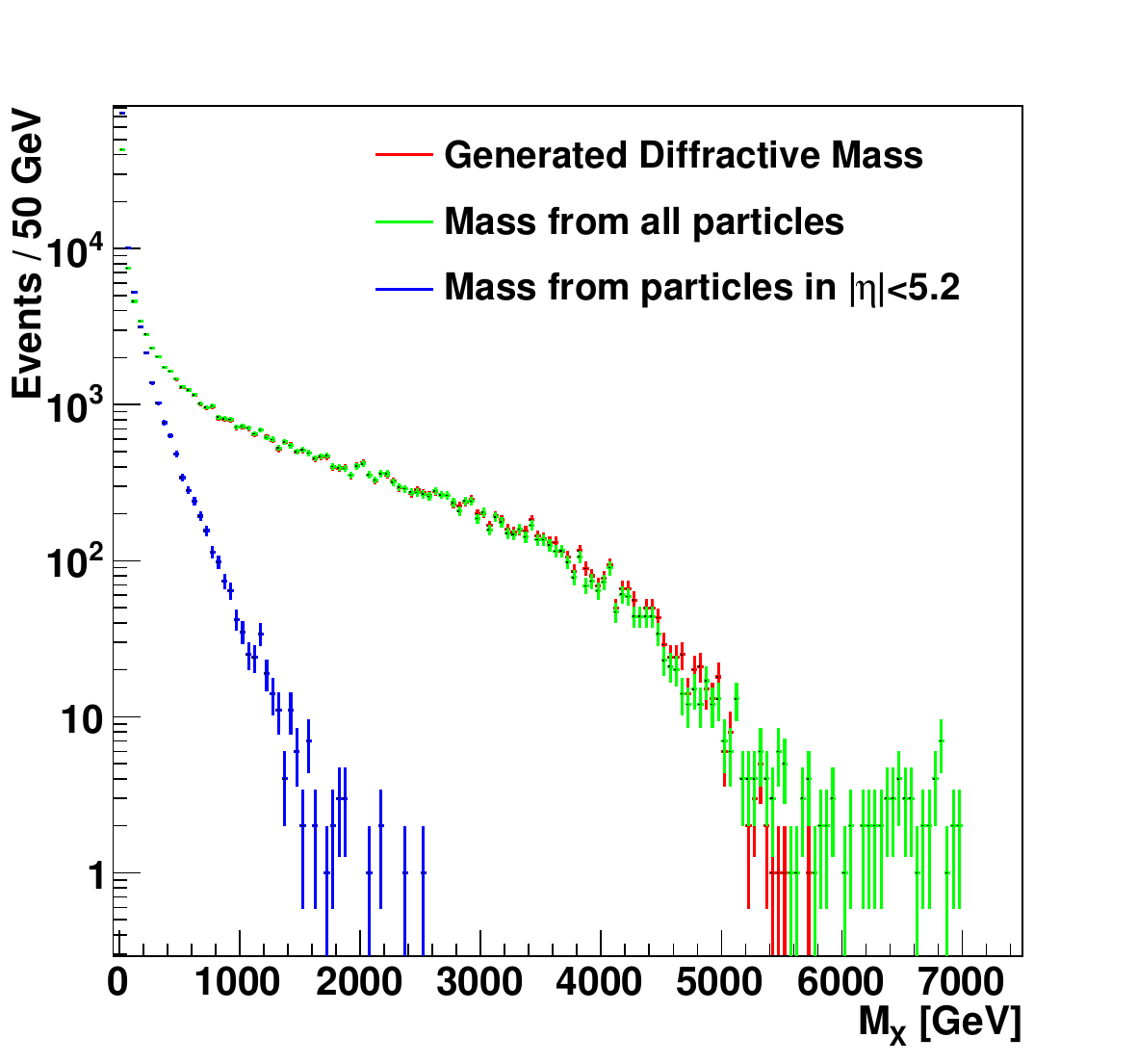}\\
    \caption{Distribution of the diffractive mass for SDD events simulated by PYTHIA 8.142.
             Generated diffractive mass (red line), calculated diffractive mass from all particles in the full $\eta$ coverage
         (green line) and in a limited $\eta$ coverage $|\eta|< 5.2$ (blue line).}
\label{fig:generatedXi}
  \end{center}
\end{figure}

The mass of the diffractive system can be measured experimentally by summing up the masses of all final 
state particles in the dissociated system as given in the following.
\begin{equation}
    M_{X} = \sum_{i}m_{i}
\end{equation}
However, it is not possible to make a precise measurement throughout the whole pseudo-rapidity range
due to the lack of the detector coverage in the very forward regions. Therefore, one can expect some difference
between the measured mass of the diffractive system and the actual mass.
This difference is illustrated in Fig.~\ref{fig:generatedXi} for single diffractive events simulated by PYTHIA 8.142~\cite{Sjostrand:2007gs}.
As shown in the figure, the diffractive mass calculated in the limited rapidity range, $|\eta|< 5.2$, does not match
the generated diffractive mass since we are not able to measure the whole mass of the
dissociated system.

It is clear that the wider the range of rapidity covered,
the more accurately the diffractive mass can be determined.
Since some of the final state particles can dissociate into the very forward rapidities, it seems not possible to measure
the actual mass of the diffractive system within the limited rapidity range $|\eta|<5.2$ which is the nominal rapidity coverage of the
CMS experiment without very forward detectors CASTOR and ZDC~\cite{Andreev:2010zzb, Grachov:2010th}.

\subsection{Large Rapidity Gap}

The gap signature in diffractive dissociation has been observed in the previous hadron-hadron collision 
experiments \cite{Abe:1994de, Aaltonen:2012tha, Acosta:2003im}.
The type of diffractive processes can be determined by looking at the number of large rapidity gaps and the position of them
in the rapidity space.
Single diffractive dissociation processes are characterised
by an edge (forward) gap only at one side of the detector while
the double diffractive dissociation processes are characterised by a central gap in the central pseudorapidity region of the detector.

The large rapidity gap in an event and the $\xi$ variable, are closely related to each other.
In SDD case, the pseudorapidity difference between the intact proton and the X system is given as $\Delta\eta$ $\sim$ -ln$\xi$.
If we experimentally measure the size of the large rapidity gap or the invariant mass of the dissociated system,
we can determine the fractional longitudinal momentum loss of the proton.

\section{Measuring Diffractive Events}

\begin{figure*}[!ht]
  \begin{center}
    \includegraphics[width=0.33\textwidth]{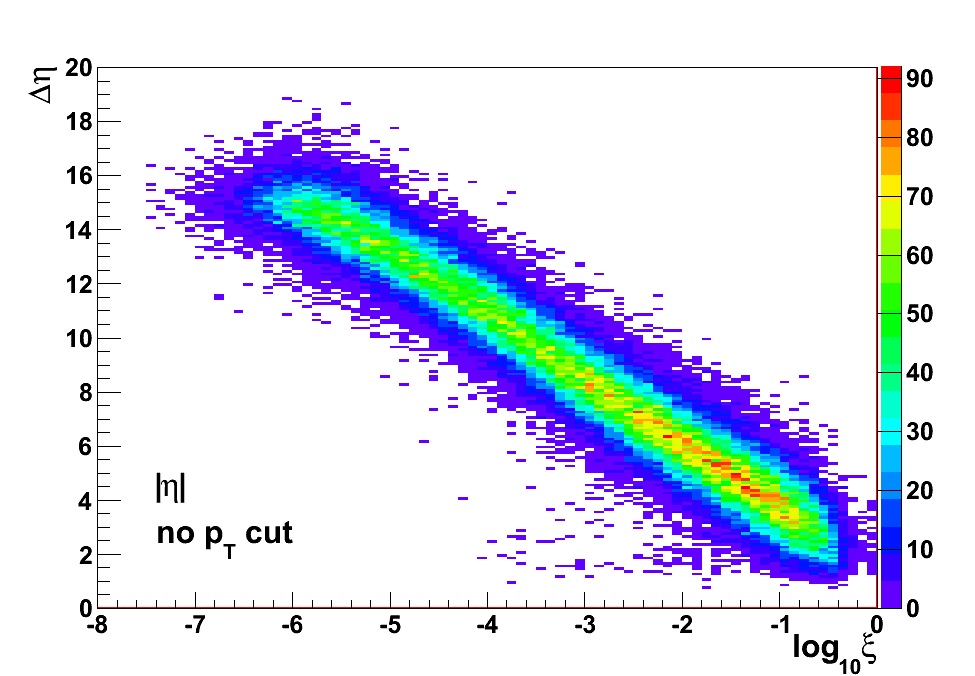}
	\includegraphics[width=0.33\textwidth]{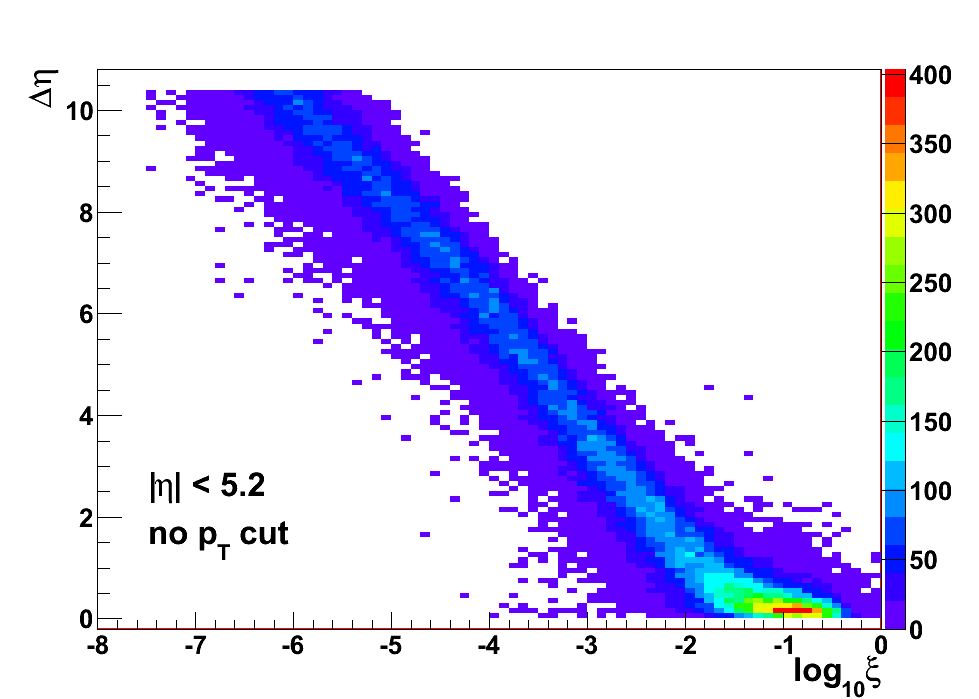}
    \includegraphics[width=0.33\textwidth]{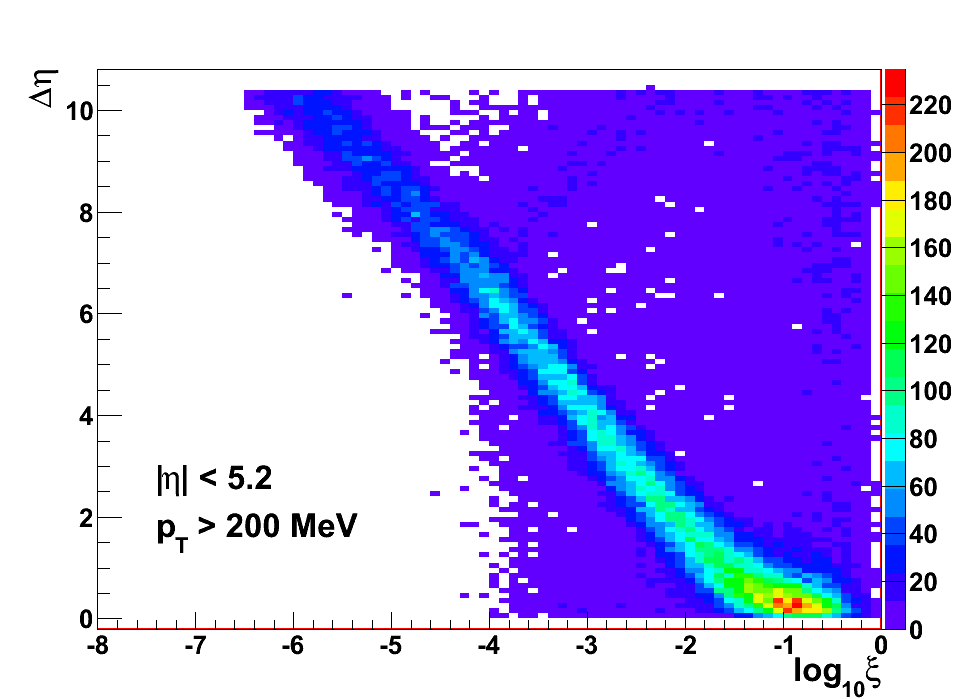}
    \caption{The relation between the size of the large rapidity gap, $\Delta\eta$, and $log_{10}\xi$
         for SDD events. The gap is defined as an edge gap.
         (a) Whole pseudo-rapidity range is used without any $p_\text{T}$ threshold on particles
         (b) For particles within $|\eta|<5.2$ without any $p_\text{T}$ threshold
         (c) For particles within $|\eta|<5.2$ with $p_{T}>$ 200 MeV.}
\label{fig:DiffMass}
  \end{center}
\end{figure*}

The measurements of the diffractive processes can be done based on the determination of the size of the large rapidity gap, $\Delta\eta$, and
the correlation between $\Delta\eta$ and $\xi$ can be used.
However, due to the forward acceptance limitations of the ATLAS and CMS detectors, it is not possible to measure
the gaps in the very forward rapidities or the whole size of the actual gap in some cases.
It is therefore important
to study the kinematical variables of the diffractive processes within the detector limits where the experimental measurements will be performed.
Fig.~\ref{fig:DiffMass} shows
the relation between the size of the large rapidity gap $\Delta\eta$, and $log_{10}\xi$
for single diffractive dissociated events.

In this paper, we use the relation $\xi=M_{X}^2/s$ for the calculation of $\xi$. First, we consider the whole pseudorapidity range and
find the largest gap in an event. All the particles with the pseudorapidity less than (or equal to) the lower
boundary of the gap are considered in one system
and the rest are in the other. Then, we add the four vectors of all particles in the given system to get the invariant mass and thus $\xi$.
The Rivet\_v1.3.0~\cite{Buckley:2010ar} analysis toolkit was used throughout the analyses in this study.

\subsection{Detector Rapidity Acceptance}

\begin{figure}[!ht]
  \begin{center}
    \includegraphics[width=0.42\textwidth]{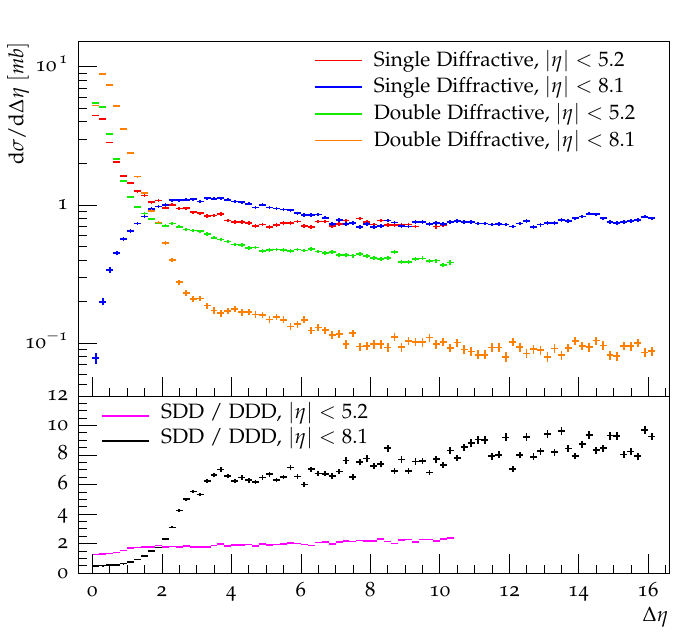}\\
    \caption{Large rapidity gap distribution for single and double diffractive dissociated events in the different
         detector rapidity acceptance. The gap is defined as an edge gap.
             Ratio of single to double diffractive dissociated events is given on the ratio plot.}
\label{fig:Acceptance}
  \end{center}
\end{figure}

We study the edge gap distribution, $\frac{d\sigma}{d\Delta\eta}$ for single and double diffractive dissociated events in the
different detector rapidity acceptance $|\eta|<5.2$ and $|\eta|<8.1$ as shown in Fig.~\ref{fig:Acceptance}.

The distributions are normalized by the cross-section of the processes obtained from PYTHIA 8. These cross-sections for different
processes are given in Table~\ref{table:sigma}. The minimum bias event class corresponds to total inelastic collisions.
As can be seen in the Fig.~\ref{fig:Acceptance}, the large rapidity gap distribution for SDD and DDD events are slightly different for the
different detector rapidity acceptance.
A clear distinction between SDD and DDD processes is possible within the larger detector acceptance $|\eta|<8.1$,
but in the limited acceptance it is not possible.
For the rest of this analysis, we use $|\eta|<5.2$ which is the rapidity coverage of the CMS experiment with its
hadron forward calorimeters located on each side of the detector.

\begin{table}[ht]
\caption{PYTHIA 8 cross-sections at $\sqrt{s}$ = 7 TeV}
\centering
\begin{tabular}{l r}
\hline
Event Class & cross-section $(\sigma[mb])$ \\
\hline
Single Diffractive Dissociation (SDD) & 13.7 \\
Double Diffractive Dissociation (DDD) & 9.3 \\
Diffractive Dissociation (SDD+DDD)   & 22.9 \\
Non Diffractive    (ND)  & 48.5 \\
\hline
Minimum Bias (SDD+DDD+ND) & 71.4 \\
\hline
\end{tabular}
\label{table:sigma}
\end{table}

\subsection{Low-$p_\text{T}$ Threshold}

\begin{figure}[!ht]
  \begin{center}
    \includegraphics[width=0.42\textwidth]{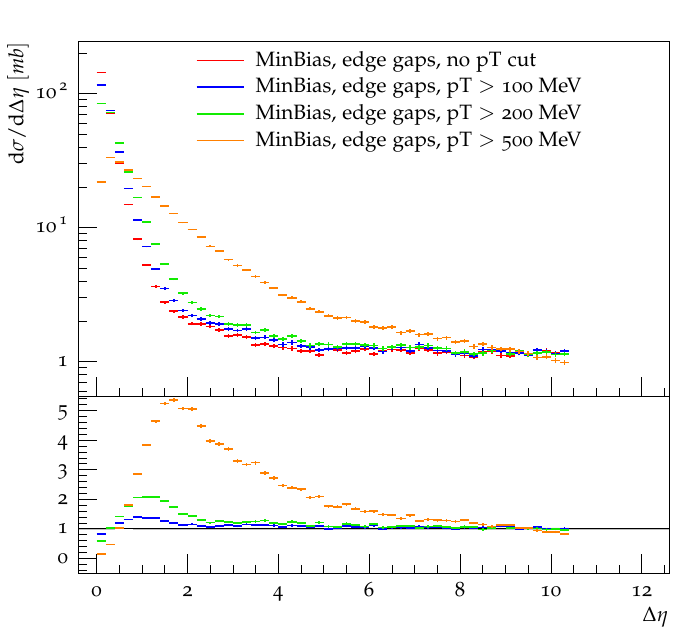}\\
    \caption{Large rapidity gap distribution for minimum bias events (SDD+DDD+ND).
         The gap is defined as an edge gap. The``MinBias, edge gaps, no pT cut'' is used as a reference
         on the ratio plot.}
\label{fig:LRGmomentum}
  \end{center}
\end{figure}

It is important to make a precise measurement of the size of the large rapidity gap, since it is directly related to the mass of the dissociated system
and the longitudinal momentum loss of the proton.
There are several factors such as radiation from multiple parton-parton
interactions, accelerator related radiation and so on, that can affect the measurement.
Also, limitations of detector response and resolution and the electronic noise will not allow
the measurement of very low-$p_\text{T}$ particles. All these factors should be considered when using the method of large rapidity gaps
for the measurement of diffractive dissociated events.
We study the different low-$p_\text{T}$ thresholds on the final state particles as represented in Fig.~\ref{fig:LRGmomentum}.
It is obvious that when the threshold is increased, some of the soft particles (i.e pions) could have lower $p_\text{T}$ than the threshold, therefore,
the gap size becomes larger.

\begin{figure*}[!ht]
  \begin{center}
    \includegraphics[width=0.44\textwidth]{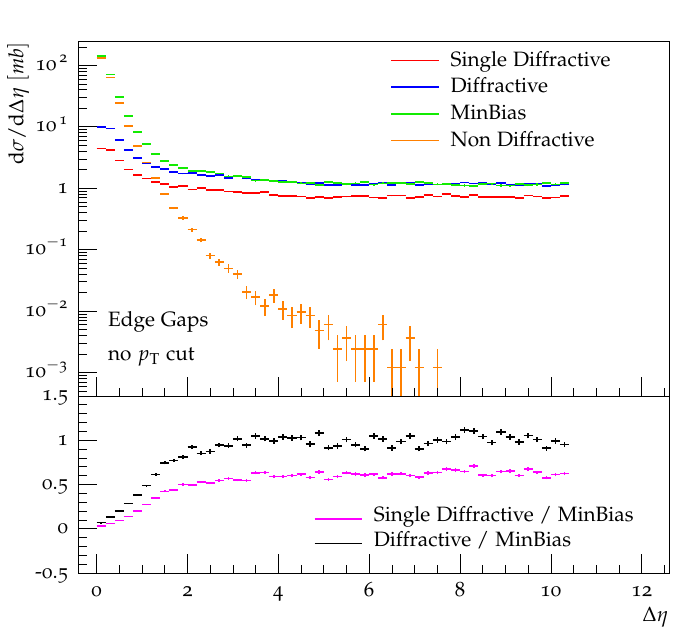}\hspace{1cm}\includegraphics[width=0.44\textwidth]{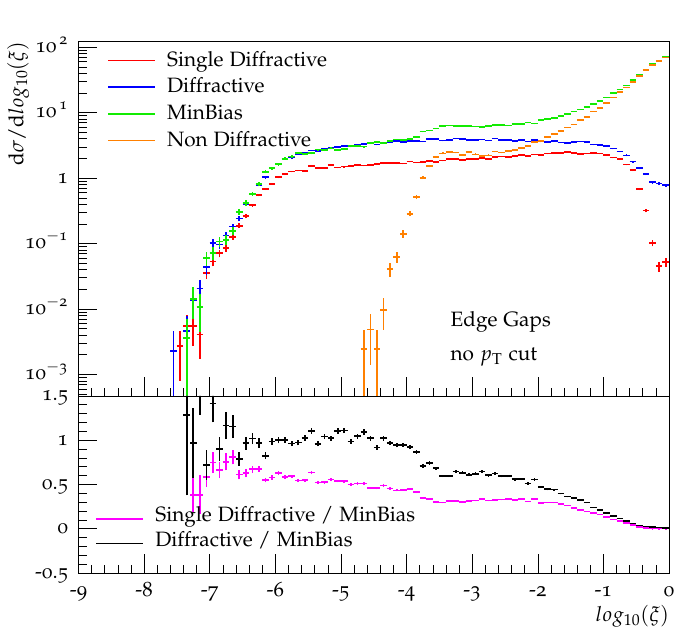}\\
    \includegraphics[width=0.44\textwidth]{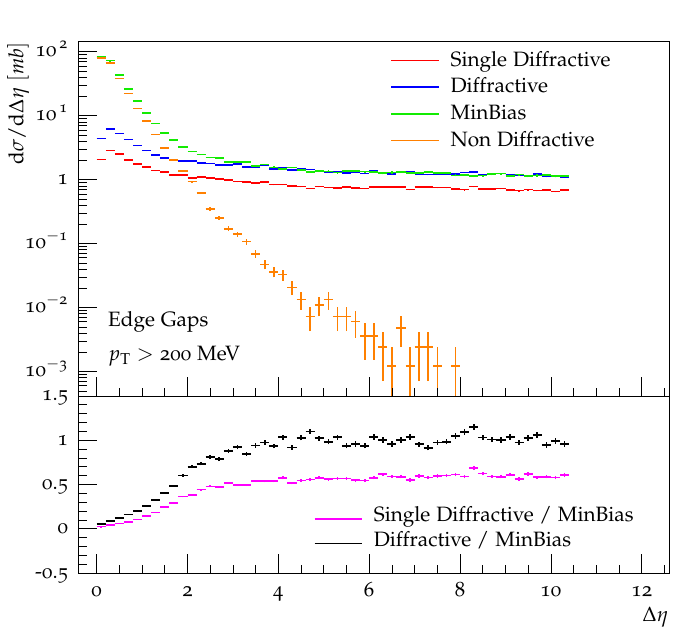}\hspace{1cm}\includegraphics[width=0.44\textwidth]{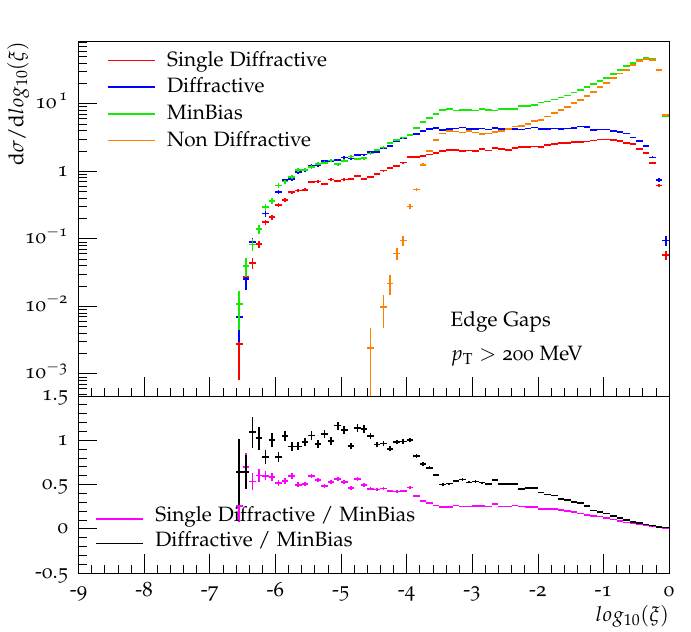}\\
    \includegraphics[width=0.44\textwidth]{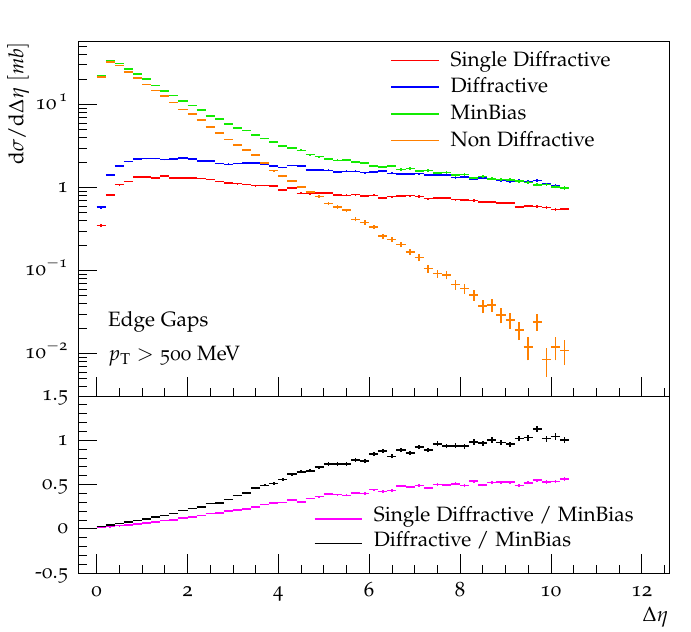}\hspace{1cm}\includegraphics[width=0.44\textwidth]{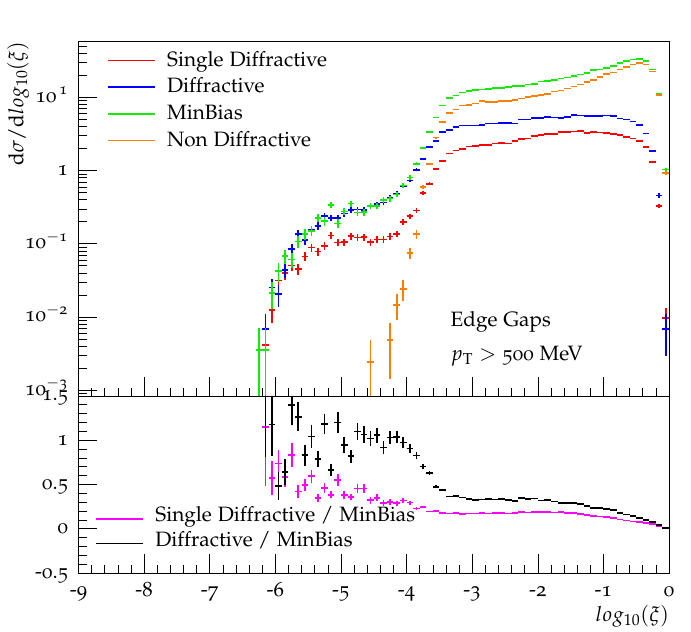}\\
    \caption{On left $\frac{d\sigma}{d\Delta\eta}$ and on right $\frac{d\sigma}{dlog_{10}\xi}$
         for different event classes. Gap is defined as edge gap and no $p_\text{T}$ cut (top),
         $p_\text{T}>$ 200 MeV (middle) and $p_\text{T}>$ 500 MeV (bottom) cuts are applied for all final state particles in $|\eta|<$ 5.2.}
\label{fig:LRGedge}
  \end{center}
\end{figure*}

\begin{figure*}[!ht]
  \begin{center}
    \includegraphics[width=0.44\textwidth]{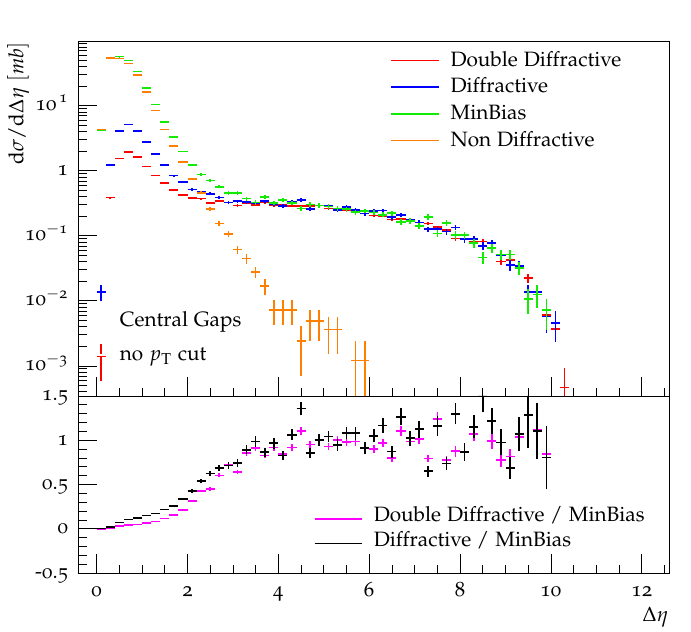}\hspace{1cm}\includegraphics[width=0.44\textwidth]{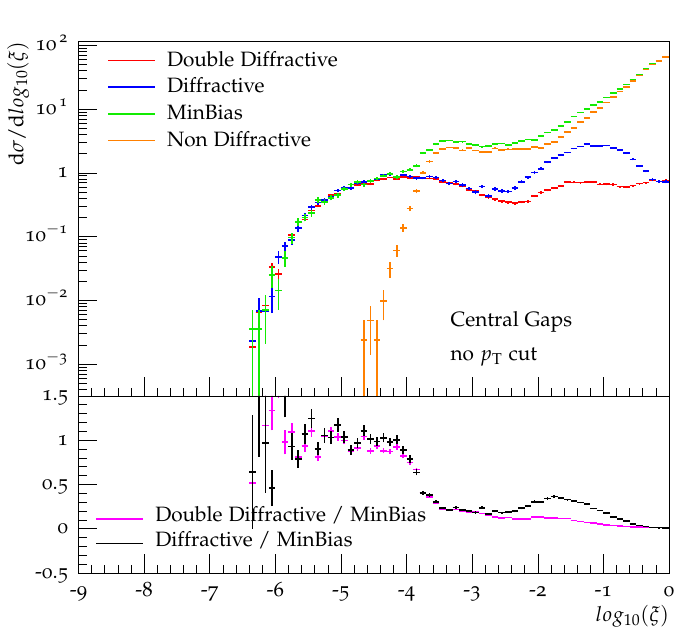}\\
    \includegraphics[width=0.44\textwidth]{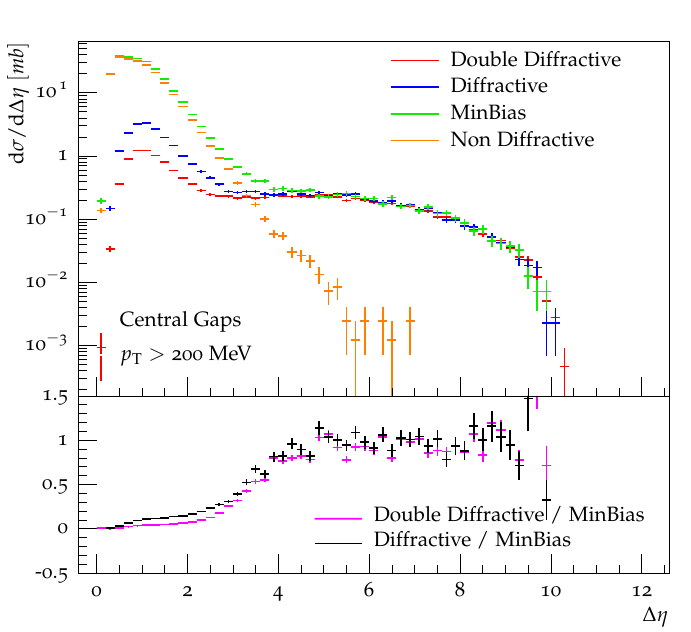}\hspace{1cm}\includegraphics[width=0.44\textwidth]{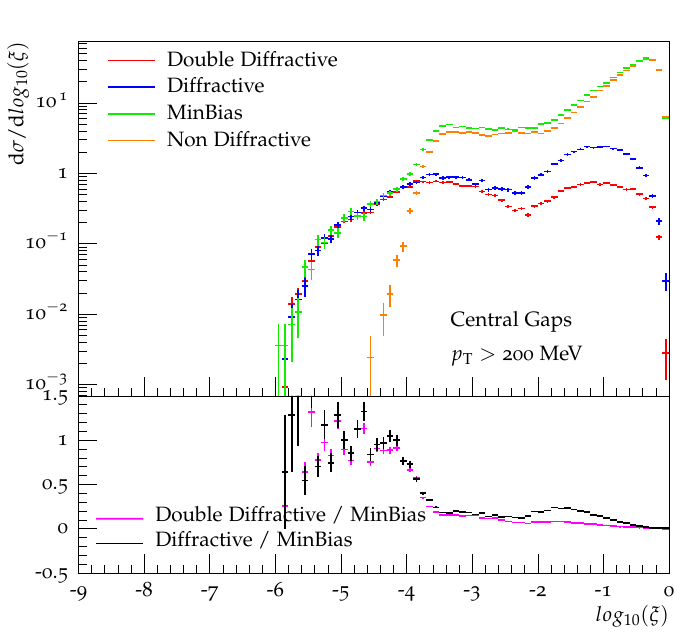}\\
    \includegraphics[width=0.44\textwidth]{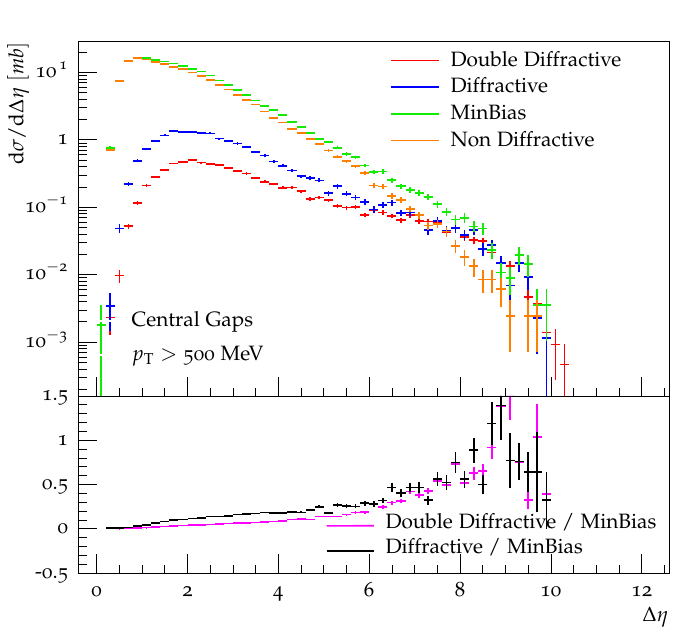}\hspace{1cm}\includegraphics[width=0.44\textwidth]{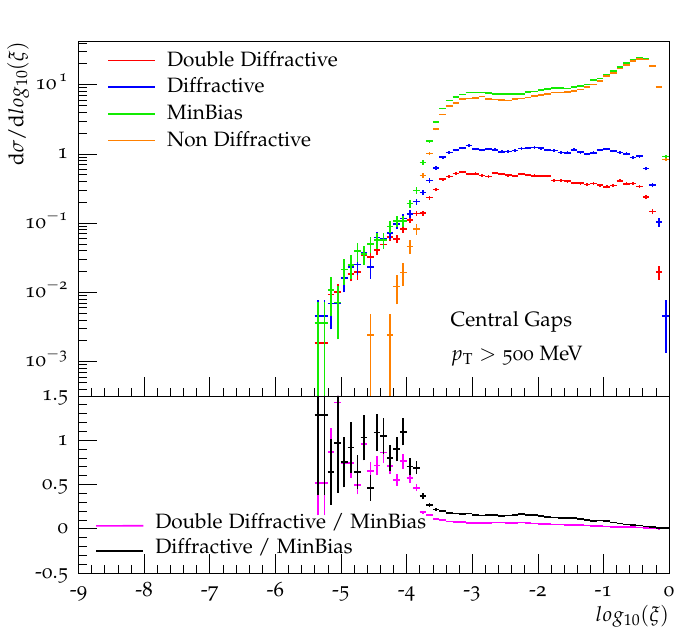}\\
    \caption{On left $\frac{d\sigma}{d\Delta\eta}$ and on right $\frac{d\sigma}{dlog_{10}\xi}$
         for different event classes. Gap is defined as central gap and no $p_\text{T}$ cut (top),
         $p_\text{T}>$ 200 MeV (middle) and $p_\text{T}>$ 500 MeV (bottom) cuts are applied for all final state particles in $|\eta|<$ 5.2.}
\label{fig:LRGcentral}
  \end{center}
\end{figure*}

In addition, the distributions of $\frac{d\sigma}{d\Delta\eta}$ and $\frac{d\sigma}{dlog_{10}\xi}$ are given
for different $p_\text{T}$ cuts in
Fig.~\ref{fig:LRGedge} for edge gaps and in Fig.~\ref{fig:LRGcentral} for central gaps.
As is clearly represented on the figures, the gap size and the ND contribution in minimum bias event content
become larger with increasing $p_\text{T}$ cut. A cut of $p_\text{T}>$ 500 MeV for all final state particles,
enhances the size of the gap for ND events.
On the other hand, each experiment must determine a reasonable $p_\text{T}$ threshold regarding the capabilities of their detector
and a low $p_\text{T}$ cut, such as 100 MeV, might not be suitable for the measurements with the data due to the detector noise at this level.
Therefore, a cut of $p_\text{T}>$ 200 MeV for all final state particles, seems to be an ideal cut to perform the measurements experimentally.
The $\Delta\eta>$ 3 cut is for edge gaps, for central gaps it looks like $\Delta\eta>$ 4 is a better cut.
When we apply $\Delta\eta>$ 3 cut with a cut of $p_\text{T}>$ 200 MeV for all final state particles in $|\eta|<$ 5.2, it seems possible to suppress
a large fraction of ND events and select the diffractive dissociated events in minimum bias data.
These cuts also will allow us to perform the measurements experimentally.

\subsection{Distinguishing SDD and DDD Events}

From a phenomenological point of view,
looking at the number and position of the large rapidity gaps in rapidity space,
one can differentiate the type of the diffractive process.
However, when we consider the detector rapidity coverage,
this may not be possible since we are not able to measure the gaps in very forward rapidities.

\subsubsection{Edge Gaps}

\begin{figure}[!ht]
  \begin{center}
    \includegraphics[width=0.42\textwidth]{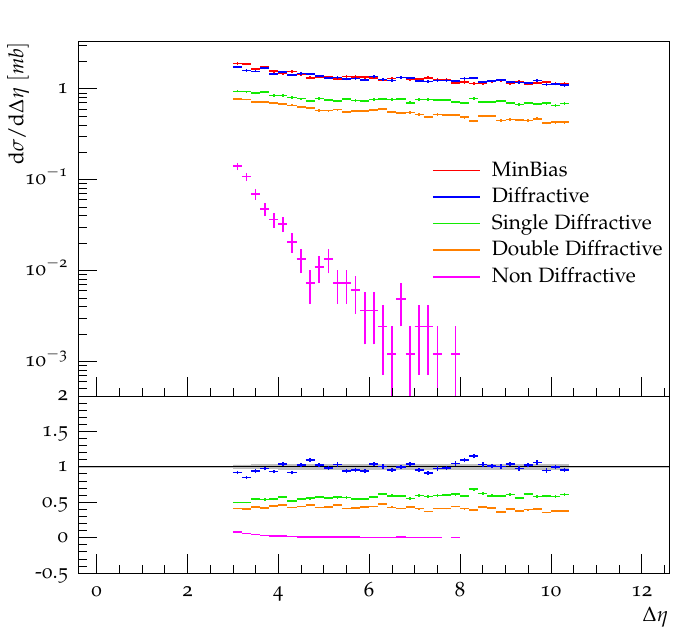}\\
    \caption{Large rapidity gap distribution for the events that have an edge gap $\Delta\eta>$ 3 with a cut of $p_\text{T}>$ 200 MeV
         for all final state particles in $|\eta|<$ 5.2. The MinBias event class is used as a reference for the ratio.}
\label{fig:LRG3pT200}
  \end{center}
\end{figure}

We study the distinguishability of SDD and DDD events for $|\eta|<5.2$ by requiring an edge gap in the events.
The visible cross-sections of events that pass the various $\Delta\eta$ cuts with $p_\text{T}>$ 200 MeV for all final state particles,
are given in Table~\ref{table:LRG}.
Similarly, the visible cross-sections for different cuts on the transverse momentum of the final state particles for
$\Delta\eta>$ 3, are given in Table~\ref{table:pT}.
The cross section for ND events is higher for high $p_\text{T}$ cuts.
On the other hand, ND events are suppressed with increasing $\Delta\eta$ cut.
If we apply a cut of $p_\text{T}>$ 200 MeV for all final state particles in $|\eta|<$ 5.2
and select the events with an edge gap $\Delta\eta>$ 3, 98.8\% of the
minimum bias events will be diffractive dissociated events. However, with these cuts one cannot distinguish SDD and DDD events since
42.1\% of the diffractive dissociated events will be DDD events.
These results are presented by a histogram in Fig.~\ref{fig:LRG3pT200}.

\begin{table}[ht]
\caption{Visible cross-sections for different $\Delta\eta$ cuts.
     A cut of $p_\text{T}>$ 200 MeV is applied for all final state particles in $|\eta|<$ 5.2
         and the gap is defined as edge gap.}
\centering
\begin{tabular}{l r r r r}
\hline
$\sigma_\text{Process}$ (mb) & $\Delta\eta>$ 2.5 & $\Delta\eta>$ 3.0 & $\Delta\eta>$ 3.5 & $\Delta\eta>$ 4.0 \\
\hline
$\sigma_\text{SDD}$       & 6.13 & 5.62 & 5.16 & 4.72 \\
$\sigma_\text{DDD}$       & 4.48 & 4.10 & 3.72 & 3.37 \\
$\sigma_\text{SDD+DDD}$   & 10.61 & 9.72 & 8.88 & 8.09 \\
$\sigma_\text{ND}$        & 0.22 & 0.11 & 0.05 & 0.03 \\ 
$\sigma_\text{MinBias}$   & 10.83 & 9.83 & 8.93 & 8.12 \\
\hline
\end{tabular}
\label{table:LRG}
\end{table}

\begin{table}[ht]
\caption{Visible cross-sections for different cuts on the transverse momentum of the final state particles in $|\eta|<$ 5.2 for $\Delta\eta>$ 3 cut.
         Gap is defined as edge gap.}
\centering
\begin{tabular}{l r r r r}
\hline
$\sigma_\text{Process}$ (mb) & no $p_\text{T}$ cut & \begin{tabular}{@{}c@{}}$p_\text{T}>$ \\ 100 MeV\end{tabular} & \begin{tabular}{@{}c@{}}$p_\text{T}>$ \\ 200 MeV\end{tabular} & \begin{tabular}{@{}c@{}}$p_\text{T}>$ \\ 500 MeV\end{tabular} \\
\hline
$\sigma_\text{SDD}$      & 5.52 & 5.60 & 5.62 & 5.80 \\
$\sigma_\text{DDD}$      & 3.42 & 3.79 & 4.10 & 5.14 \\
$\sigma_\text{SDD+DDD}$  & 8.94 & 9.39 & 9.72 & 10.94 \\
$\sigma_\text{ND}$       & 0.03 & 0.06 & 0.11 & 4.38 \\
$\sigma_\text{MinBias}$  & 8.97 & 9.45 & 9.83 & 15.32 \\
\hline
\end{tabular}
\label{table:pT}
\end{table}

\subsubsection{Central Gaps}

\begin{figure}[!ht]
  \begin{center}
    \includegraphics[width=0.42\textwidth]{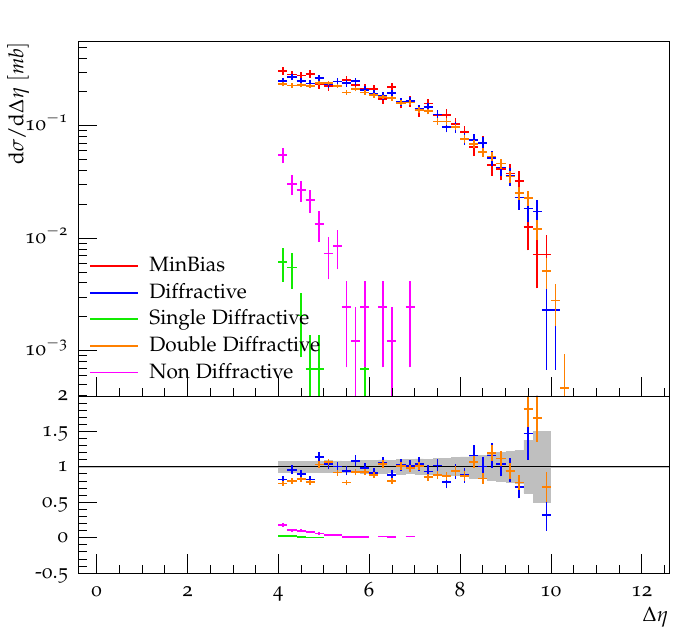}\\
    \caption{Large rapidity gap distribution for the events that have a central gap $\Delta\eta>$ 4 with a cut of $p_\text{T}>$ 200 MeV
             for all final state particles in $|\eta|<$ 5.2. The MinBias event class is used as a reference for the ratio.}
\label{fig:LRG3pT200central}
  \end{center}
\end{figure}

We perform a similar event selection by requiring a central gap in the events.
These events should be dominated by DDD events where both diffraction systems are in the calorimeter acceptance.
The visible cross-sections of events that pass the various $\Delta\eta$ cuts with $p_\text{T}>$ 200 MeV
for all final state particles are given in Table~\ref{table:LRGcentral}.
The visible cross-section for different cuts on the transverse momentum of the final state particles for
$\Delta\eta>$ 4, are presented in Table~\ref{table:pTcentral}.
As indicated in the tables, the visible cross-section decreases with increasing $\Delta\eta$.
The ND contribution in minimum bias data is dominant in the larger
$p_\text{T}$ and also, with the increasing $p_\text{T}$ cut, visible cross-section for SDD events increases while it decreases for DDD events.
If we apply a cut of $p_\text{T}>$ 200 MeV for all final state particles in $|\eta|<$ 5.2
and select the events with a central gap $\Delta\eta>$ 4, 95.3\% of the minimum bias event
will be diffractive dissociated events. SDD events in this diffractive dissociated event content
will be almost completely suppressed.
These results are summarized by a histogram in Fig.~\ref{fig:LRG3pT200central}.

\begin{table}[ht]
\caption{Visible cross-sections for different $\Delta\eta$ cuts.
         A cut of $p_\text{T}>$ 200 MeV is applied for all final state particles in $|\eta|<$ 5.2
         and the gap is defined as central gap.}
\centering
\begin{tabular}{l r r r r}
\hline
$\sigma_\text{Process}$ (mb) & $\Delta\eta>$ 2.5 & $\Delta\eta>$ 3.0 & $\Delta\eta>$ 3.5 & $\Delta\eta>$ 4.0 \\
\hline
$\sigma_\text{SDD}$      & 0.08 & 0.03 & 0.01 & 0.003 \\
$\sigma_\text{DDD}$      & 1.15 & 1.04 & 0.93 & 0.82 \\
$\sigma_\text{SDD+DDD}$  & 1.23 & 1.07 & 0.94 & 0.82 \\
$\sigma_\text{ND}$       & 0.66 & 0.22 & 0.08 & 0.04 \\
$\sigma_\text{MinBias}$  & 1.89 & 1.29 & 1.02 & 0.86 \\
\hline
\end{tabular}
\label{table:LRGcentral}
\end{table}

\begin{table}[ht]
\caption{Visible cross-sections for different cuts on the transverse momentum of the final state particles in $|\eta|<$ 5.2 for $\Delta\eta>$ 4 cut.
         Gap is defined as central gap.}
\centering
\begin{tabular}{l r r r r}
\hline
$\sigma_\text{Process}$ (mb) & no $p_\text{T}$ cut & \begin{tabular}{@{}c@{}}$p_\text{T}>$ \\ 100 MeV\end{tabular} & \begin{tabular}{@{}c@{}}$p_\text{T}>$ \\ 200 MeV\end{tabular} & \begin{tabular}{@{}c@{}}$p_\text{T}>$ \\ 500 MeV\end{tabular} \\
\hline
$\sigma_\text{SDD}$      & 0.002 & 0.001 & 0.003 & 0.18 \\
$\sigma_\text{DDD}$      & 0.95  & 0.92  & 0.82  & 0.44 \\
$\sigma_\text{SDD+DDD}$  & 0.95  & 0.92  & 0.82  & 0.62 \\
$\sigma_\text{ND}$       & 0.007 & 0.01  & 0.04  & 1.99 \\
$\sigma_\text{MinBias}$  & 0.96  & 0.93  & 0.86  & 2.61 \\
\hline
\end{tabular}
\label{table:pTcentral}
\end{table}

Although central gaps look like an ideal cut to separate SDD and DDD events, only a small fraction of the DDD cross-section has
a central gap. It mostly has an edge gap. Due to a class of DDD events with a low diffractive mass on one side, the particles
beyond the acceptance of the detector are not detected and they look like SDD events in the limited detector rapidity acceptance.
We calculate the fraction of DDD events which can be tagged as SDD events in the limited rapidity range $|\eta|<5.2$.
Results for different cuts on the size of the edge gap $\Delta\eta$, are given in Table~\ref{table:ZDC2}.
As indicated in the table, for $\Delta\eta>$ 3, 44.1\% of the DDD
events can be tagged as SDD events in the limited detector rapidity coverage. This event fraction is smaller for the larger gap sizes.

\begin{table}[ht]
\caption{The fraction of DDD events which can be tagged as SDD events in the limited rapidity range $|\eta|<5.2$.
     Gap is defined as edge gap.}
\centering
\begin{tabular}{l r r r r}
\hline
Event Class & $\Delta\eta>$ 2.5 & $\Delta\eta>$ 3.0 & $\Delta\eta>$ 3.5 & $\Delta\eta>$ 4.0 \\
\hline
DDD & 48.2\% & 44.1\% & 40.2\% & 36.3\% \\
\hline
\end{tabular}
\label{table:ZDC2}
\end{table}

\subsubsection{Multiplicity and Total Energy Deposition}

\begin{figure*}[!ht]
  \begin{center}
    \includegraphics[width=0.33\textwidth]{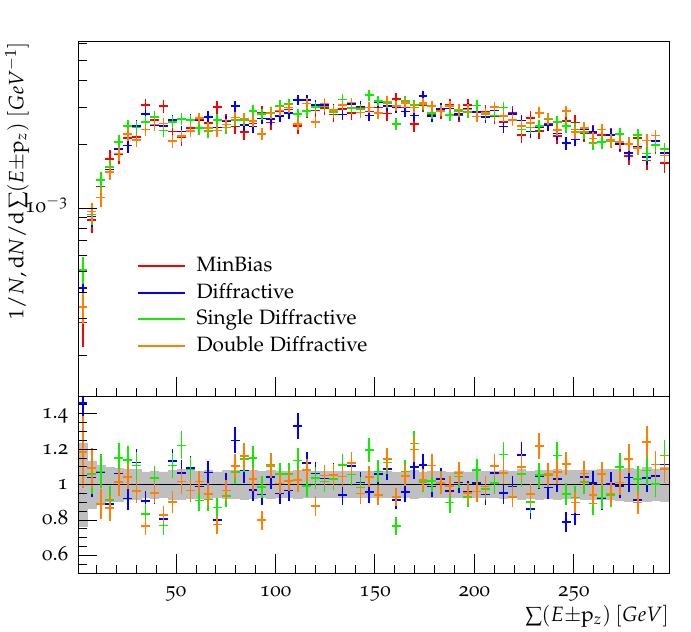}
	\includegraphics[width=0.33\textwidth]{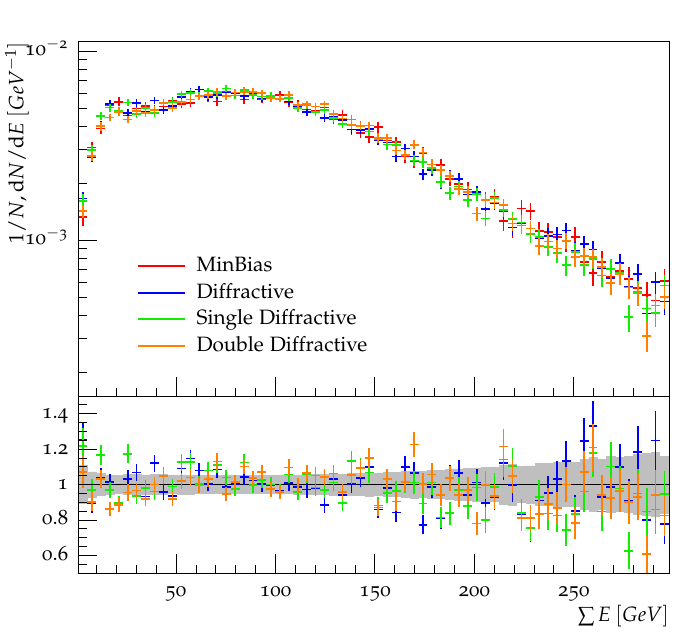}
    \includegraphics[width=0.33\textwidth]{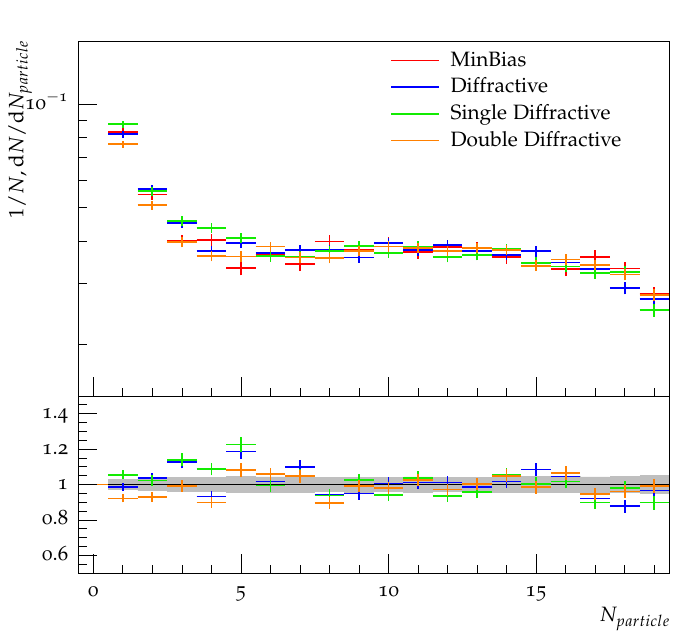}
    \caption{Distributions of $\sum$$(E\pm$ $p_{z})$ (left), total energy deposition (right) and particle multiplicity (bottom)
         for the events that have an edge gap $\Delta\eta>$ 3 with a $p_\text{T}>$ 200 MeV cut on final state particles in $|\eta|<$ 5.2.
         The distributions are normalized to the number of events that pass the analysis cuts. The MinBias event class is used as a reference
         for the ratio.}
\label{fig:commondist}
  \end{center}
\end{figure*}

In order to try and distinguish SDD events
from DDD events with a low diffractive mass, the distributions $\sum$($E\pm$$p_{z}$),
total energy deposition and particle multiplicity were investigated.
The sum for $\sum$($E\pm$$p_{z}$), runs over all final state particles in $|\eta|<5.2$. The longitudinal momentum $p_{z}$ is calculated
as $Ecos\theta$, where $E$ is the energy of the particle and $\theta$ is the angle between the particle momentum $\vec{p}$
and the beam axis. The distributions for the events which have an edge gap $\Delta\eta>$ 3 with a cut of $p_\text{T}>$ 200 MeV for all final state particles,
are given in Fig.~\ref{fig:commondist}. As shown in the figure, the shape of the distributions for different event classes look very similar
and therefore, it seems not possible to separate SDD from DDD events with these cuts by using edge gaps.
The distributions for ND events class
are not represented since there are very few ND events which pass the event selection.

\begin{figure*}[!ht]
  \begin{center}
    \includegraphics[width=0.33\textwidth]{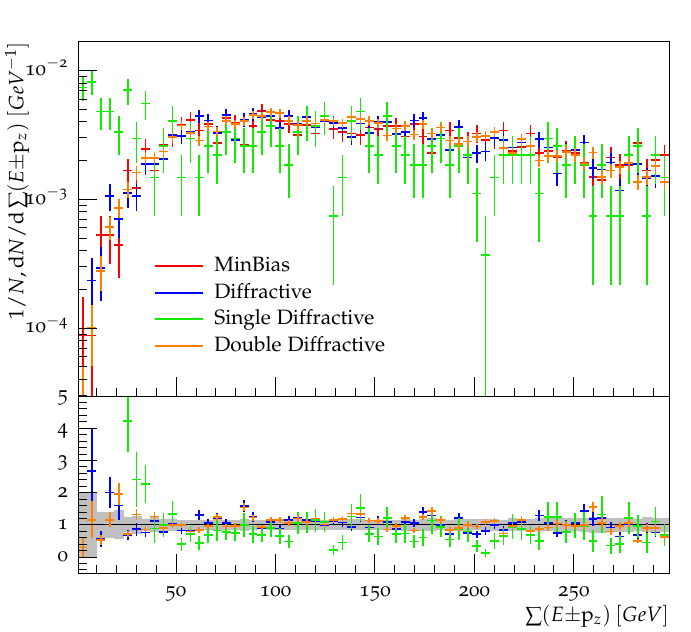}
	\includegraphics[width=0.33\textwidth]{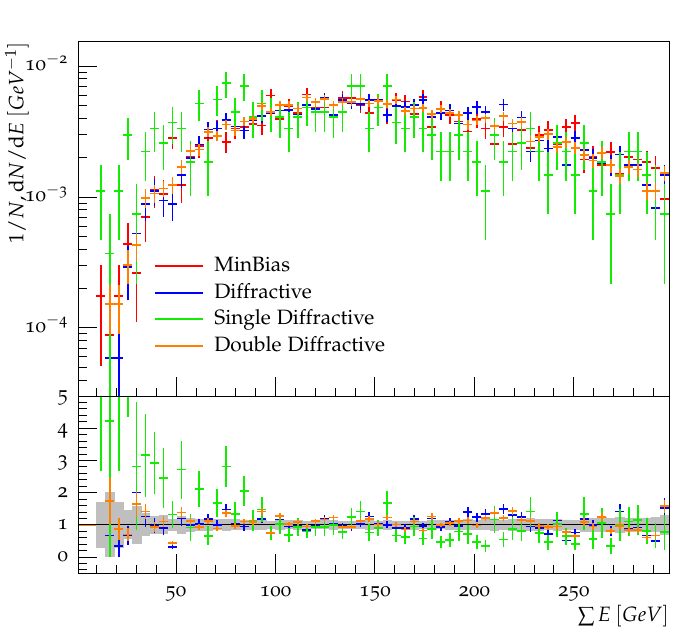}
    \includegraphics[width=0.33\textwidth]{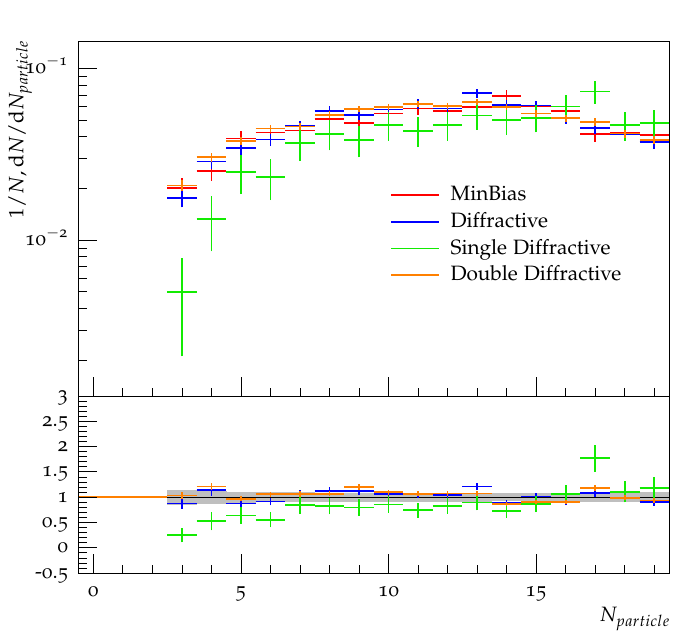}
    \caption{Distributions of $\sum$$(E\pm$ $p_{z})$ (left), total energy deposition (right) and particle multiplicity (bottom)
         for the events that have a central gap $\Delta\eta>$ 4 with a $p_\text{T}>$ 200 MeV cut on final state particles in $|\eta|<$ 5.2.
         The distributions are normalized to the number of events that pass the analysis cuts. The MinBias event class is used as a reference
         for the ratio.}
\label{fig:commondistCENtral}
  \end{center}
\end{figure*}

The same distributions were studied also for central gaps.
The distributions for the events which have a central gap $\Delta\eta>$ 4 with a cut of $p_\text{T}>$ 200 MeV for all final state particles in $|\eta|<5.2$,
are given in Fig.~\ref{fig:commondistCENtral}.
The distributions for events with an edge gap do not distinguish SDD and DDD (Fig.~\ref{fig:commondist}),
and that for central gaps (Fig.~\ref{fig:commondistCENtral}) there are some differences, but the SDD contribution is anyway very suppressed in these events.

\subsubsection{SDD and DDD Events at Very Forward Rapidities}

As we already discussed in the previous sections, a class of DDD events with a low diffractive mass on one side can be tagged
as SDD events in the limited rapidity acceptance of the detector. Since the particles in such events dissociate into the forward rapidities,
looking at the particle activity in the very forward detectors can provide more accurate information about the type of the process.
The ATLAS and CMS Zero Degree Calorimeters, ZDC, are located $\sim$140 meters away from the interaction point (IP) on both sides,
at the end of the straight LHC beam-line section.
The ZDC cover the psedudorapidity region $|\eta|>8.1$ and are able to detect very forward
neutral particles ($n, \gamma$, $\pi^{\circ}$) at a $0^{\circ}$ polar angle.

\begin{figure*}[!ht]
  \begin{center}
    \includegraphics[width=0.42\textwidth]{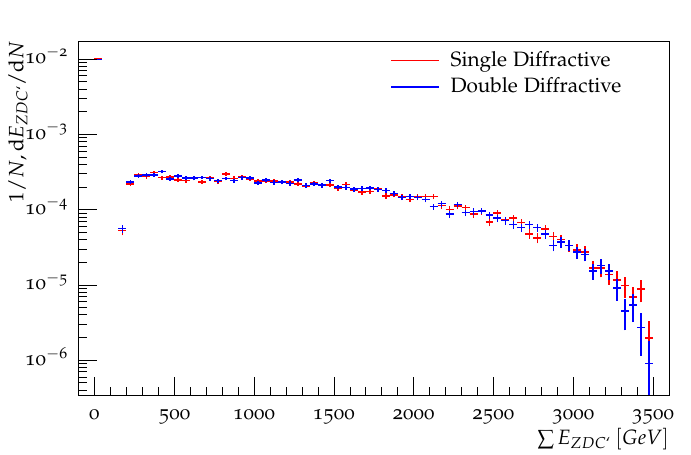}\hspace{0.5cm}
	\includegraphics[width=0.42\textwidth]{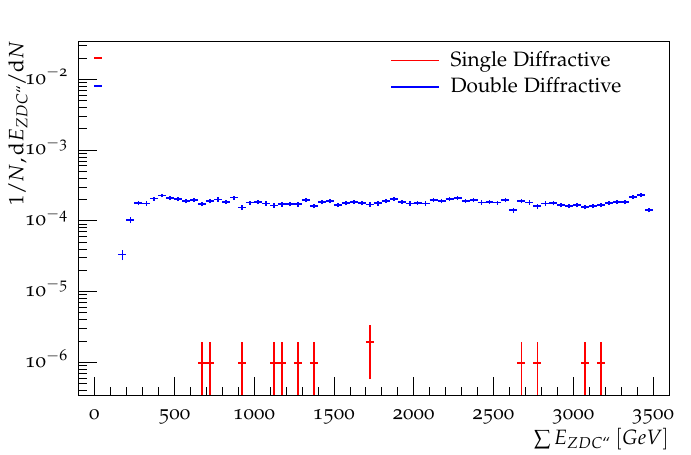}\\
	(a)      \hspace{7.5cm}   (b)\\
    \includegraphics[width=0.42\textwidth]{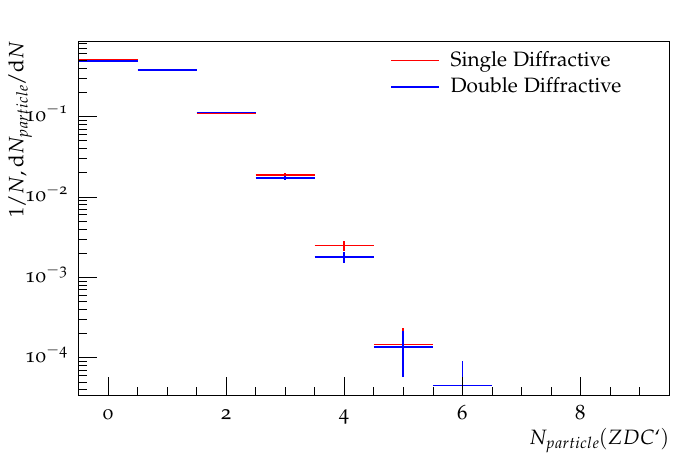}\hspace{0.5cm}
	\includegraphics[width=0.42\textwidth]{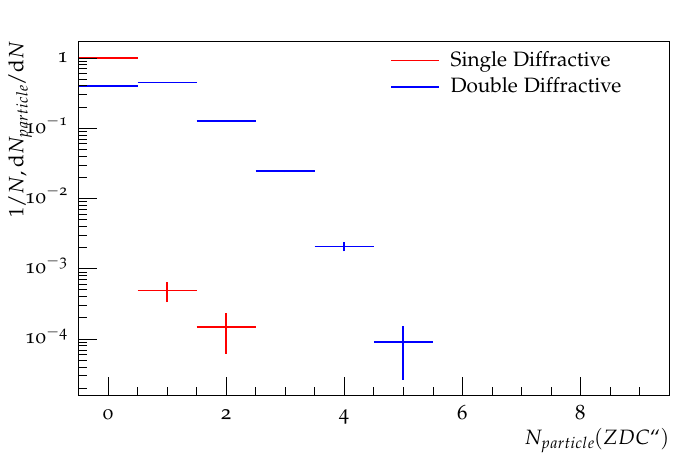}\\
	(c)      \hspace{7.5cm}   (d)
    \caption{The total energy deposition and the multiplicity of the neutral particles with $p_\text{T}^{0}>$ 100 MeV
         in the ZDC detectors for the events that have an edge gap in $|\eta|<5.2$
         with a gap size $\Delta\eta>$ 3, and with $E\neq0$ in the opposite side from the gap (either at $\eta<$ 0 or $\eta>$ 0 depending on the
         gap position).
         A cut of $p_\text{T}>$ 200 MeV was applied for the final state particles in $|\eta|<5.2$ to find the size of the gap.
         ZDC$'$ refers to the ZDC which is on the side opposite the gap and ZDC$''$ is the ZDC on the side with the gap.
         Total energy deposition in (a) ZDC$'$ and (b) ZDC$''$
         and particle multiplicity in (c) ZDC$'$ and (d) ZDC$''$ are given.
         The distributions are normalized to the number of events that pass the analysis cuts.}
\label{fig:ZDCplot}
  \end{center}
\end{figure*}

We look at the total energy deposition and the multiplicity of the neutral particles in the ZDC acceptance in order
to investigate whether there is a way to separate SDD and low-mass DDD events.
An edge gap was required in the events with a gap size $\Delta\eta>$ 3 and with a cut of
$p_\text{T}>$ 200 MeV for all final state particles in $|\eta|<$ 5.2.
Additionally, a certain amount of energy deposition $E\neq0$ was required in the opposite side from the gap
(either at $\eta<$ 0 or $\eta>$ 0 depending on the gap position).
In Fig. \ref{fig:ZDCplot},
the total energy deposition and the multiplicity of the neutral particles in
ZDC$'$ (the ZDC detector on the side opposite the gap) and ZDC$''$ (the ZDC detector on the side with the gap) is given
with a cut of 100 MeV for the transverse momentum of the neutral particles ($p_\text{T}^{0}$).
The $p_\text{T}^{0}>$ 100 MeV is a reasonable cut given the ZDC noise levels.
Also, the fraction of the events which have at least one neutral particle in ZDC$''$
for different cuts on the transverse momentum of the neutral particles with
different cuts on the size of the gap, is given in Table~\ref{table:ZDCSDD} for SDD and in Table~\ref{table:ZDCDDD} for DDD processes.
As can be seen, for the events that have an edge gap with a size of $\Delta\eta>$ 3 and $p_\text{T}^{0}>$ 100 MeV,
SDD events have almost no neutral particle in ZDC$''$ while 60.2\% of the DDD events have at least one neutral particle with $p_\text{T}^{0}>$ 100 MeV
scattering into these forward rapidities.
Although ZDC can provide a better distinction for SDD and DDD events, about 60\% of the DDD events do not have any
particles with in ZDC$''$
and it is therefore, they cannot be distinguished from SDD events in $|\eta|<5.2$. Looking at the total energy deposition and
the particle multiplicity in the ZDC detectors is not
enough to select a pure sample of SDD or DDD events. For the separation of SDD and low-mass DDD events,
we must perform the measurements in the whole range of the rapidity space.

\begin{table}[ht]
\caption{The fraction of SDD events which have at least one neutral particle in ZDC$''$ (the ZDC detector on the side with the gap),
     is given for different cuts on the size of the $\Delta\eta$ with different thresholds for the transverse momentum of the neutral particles.
     The gap is defined as an edge gap and the final state particles within $|\eta|<5.2$ with $p_T>$ 200 MeV are used to find the size of the gap.}
\centering
\begin{tabular}{l r r r r}
\hline
& $\Delta\eta>$ 2.5 & $\Delta\eta>$ 3.0 & $\Delta\eta>$ 3.5 & $\Delta\eta>$ 4.0 \\
\hline
no $p_\text{T}^{0}$ cut & 0.23\% & 0.10\% & 0.06\% & 0.03\% \\
$p_\text{T}^{0}>$ 100 MeV & 0.20\% & 0.09\% & 0.05\% & 0.03\% \\
$p_\text{T}^{0}>$ 200 MeV & 0.16\% & 0.08\% & 0.05\% & 0.02\% \\
\hline
\end{tabular}
\label{table:ZDCSDD}
\end{table}

\begin{table}[ht]
\caption{The fraction of DDD events which have at least one neutral particle in ZDC$''$ (the ZDC detector on the side with the gap),
     is given for different cuts on the size of the $\Delta\eta$ with different thresholds for the transverse momentum of the neutral particles.
     The gap is defined as an edge gap and the final state particles within $|\eta|<5.2$ with $p_{T}>$ 200 MeV are used to find the size of the gap.}
\centering
\begin{tabular}{l r r r r}
\hline
& $\Delta\eta>$ 2.5 & $\Delta\eta>$ 3.0 & $\Delta\eta>$ 3.5 & $\Delta\eta>$ 4.0 \\
\hline
no $p_\text{T}^{0}$ cut & 68.93\% & 68.90\% & 68.93\% & 68.97\% \\
$p_\text{T}^{0}>$ 100 MeV & 60.28\% & 60.20\% & 60.18\% & 60.22\% \\
$p_\text{T}^{0}>$ 200 MeV & 47.27\% & 47.09\% & 46.89\% & 46.76\% \\
\hline
\end{tabular}
\label{table:ZDCDDD}
\end{table}

\subsection{Bias From Vertex}

One of the background sources of diffractive processes at the LHC is the radiation coming from
non-colliding bunches. The common practice to eliminate such background is usually to
require a primary vertex in the events. The primary vertex is defined as the location of $pp$ collision. 
The number of tracks, or theoretically number of charged particles, associated with a primary vertex can be different for different processes.
In some cases, such as low-mass diffractive dissociation, the system can dissociate into the very forward rapidities and thus
all the particles may appear at small polar angles. The problem in this case is the limited detector instruments in this region.
The tracker of ATLAS and CMS, which is used to measure charged particles, covers the pseudorapidity region $|\eta| < 2.5$.
Therefore, it will not be possible to measure charged particles when the system dissociates into the very forward
rapidities. These types of events may not form a reconstructable primary vertex if all the particles are outside of the tracker region.
We study the bias on the gap distribution introduced by requiring a vertex.

\begin{figure}[!ht]
  \begin{center}
    \includegraphics[width=0.42\textwidth]{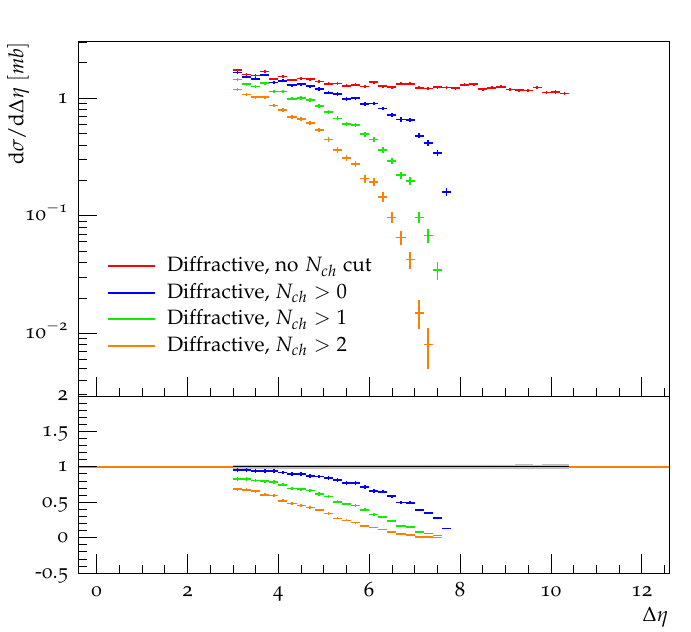}\\
    \caption{Diffractive dissociated events without a primary vertex and with a primary vertex for the different number of charged particles,
         $N_\text{ch}$.
         Tracker region is considered as $|\eta| < 2.5$, and $p_\text{T}>$ 200 MeV charged particles are used to form the vertex.
             Only the events that have an edge gap $\Delta\eta>$ 3 with a cut of $p_\text{T}>$ 200 MeV for all final state particles
         in $|\eta|<$ 5.2 are considered. The vertex requirement suppresses the events that have $\Delta\eta>$ 8 which
         corresponds to the very low-mass soft diffractive processes.}
\label{fig:Vertex}
  \end{center}
\end{figure}

Fig.~\ref{fig:Vertex} shows the large rapidity gap distribution for diffractive events with and without a primary vertex
where the tracker region is considered as $|\eta| < 2.5$ and $p_\text{T}>$ 200 MeV charged particles are used to form the vertex.
The fraction of events that pass the cut, for different number of charged particles, is given in Table~\ref{table:Tracks} for
different event classes. Requiring a primary vertex which is reconstructed with two or more charged particles, suppress at least 33.2\% of
the diffractive events. Particularly, very low-mass soft diffractive dissociated events with a very large gap, $\Delta\eta>$ 8, are suppressed by
the primary vertex cut.
Therefore, instead of a primary vertex cut one should investigate other experimental ways of eliminating the background
coming from non-colliding bunches. In this limited tracker range, a primary vertex cut is not practical to perform measurements for
the diffractive dissociated events.

\begin{table}[ht]
\caption{The fraction of events without a primary vertex, and with a primary vertex for the different number of charged particles,
     $N_\text{ch}$. Tracker region is considered as $|\eta| < 2.5$ and $p_\text{T}>$ 200 MeV charged particles are used to form the vertex.
     Only the events
     that have an edge gap $\Delta\eta>$ 3 with a cut of $p_\text{T}>$ 200 MeV for all final state particles in $|\eta|<$ 5.2 are considered.}
\centering
\begin{tabular}{l c c c c}
\hline
Event Class & no $N_\text{ch}$ cut & $N_\text{ch}>$ 0 & $N_\text{ch}>$ 1 & $N_\text{ch}>$ 2 \\
\hline
Single Diffractive & 41.1\% & 19.9\% & 13.1\% & 8.5\% \\
Double Diffractive & 44.3\% & 23.0\% & 15.4\% & 10.1\% \\
Diffractive    & 42.4\% & 21.1\% & 14.1\% & 9.2\% \\
MinBias        & 13.8\% & 7.0\% & 4.7\% & 3.1\% \\
Non Diffractive    & 0.2\%  & 0.2\% & 0.2\% & 0.1\% \\
\hline
\end{tabular}
\label{table:Tracks}
\end{table}

\subsection{Model Dependency}

\begin{figure}[h!t]
  \begin{center}
    \includegraphics[width=0.44\textwidth]{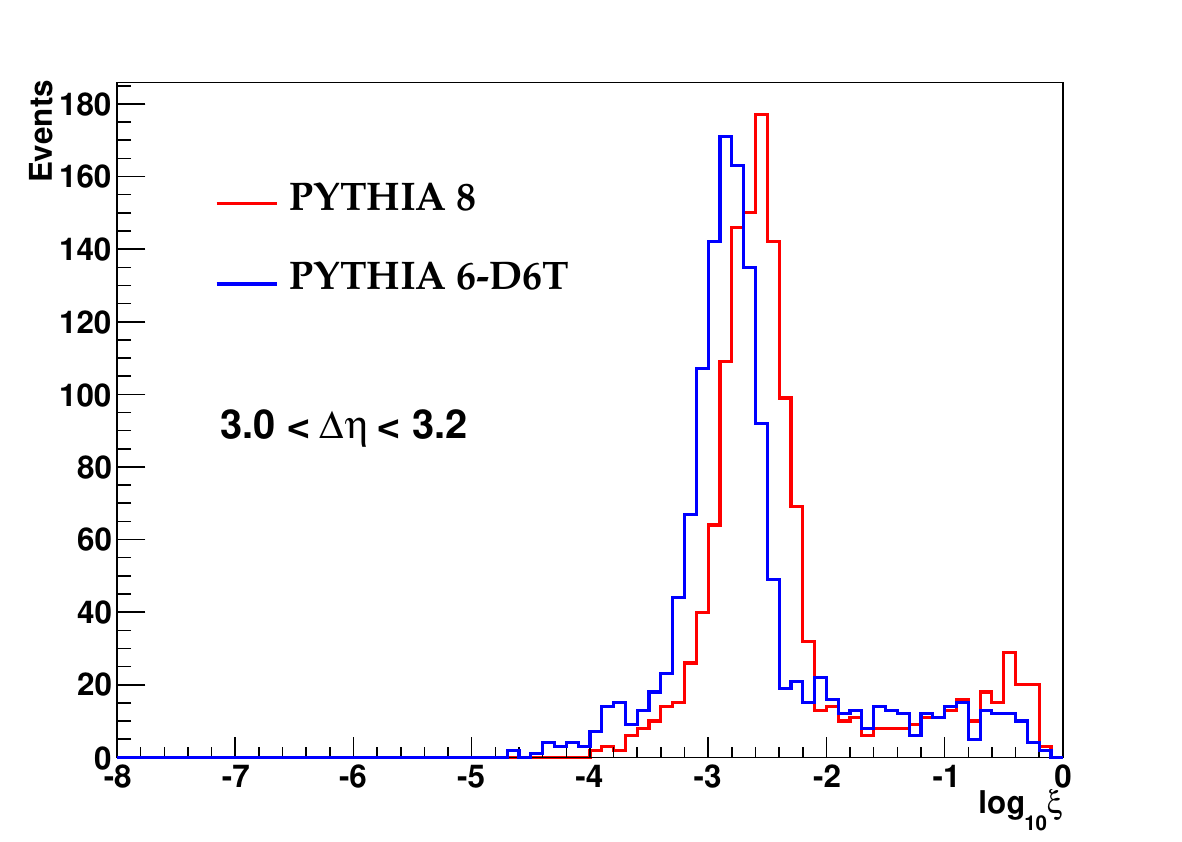}
	\includegraphics[width=0.44\textwidth]{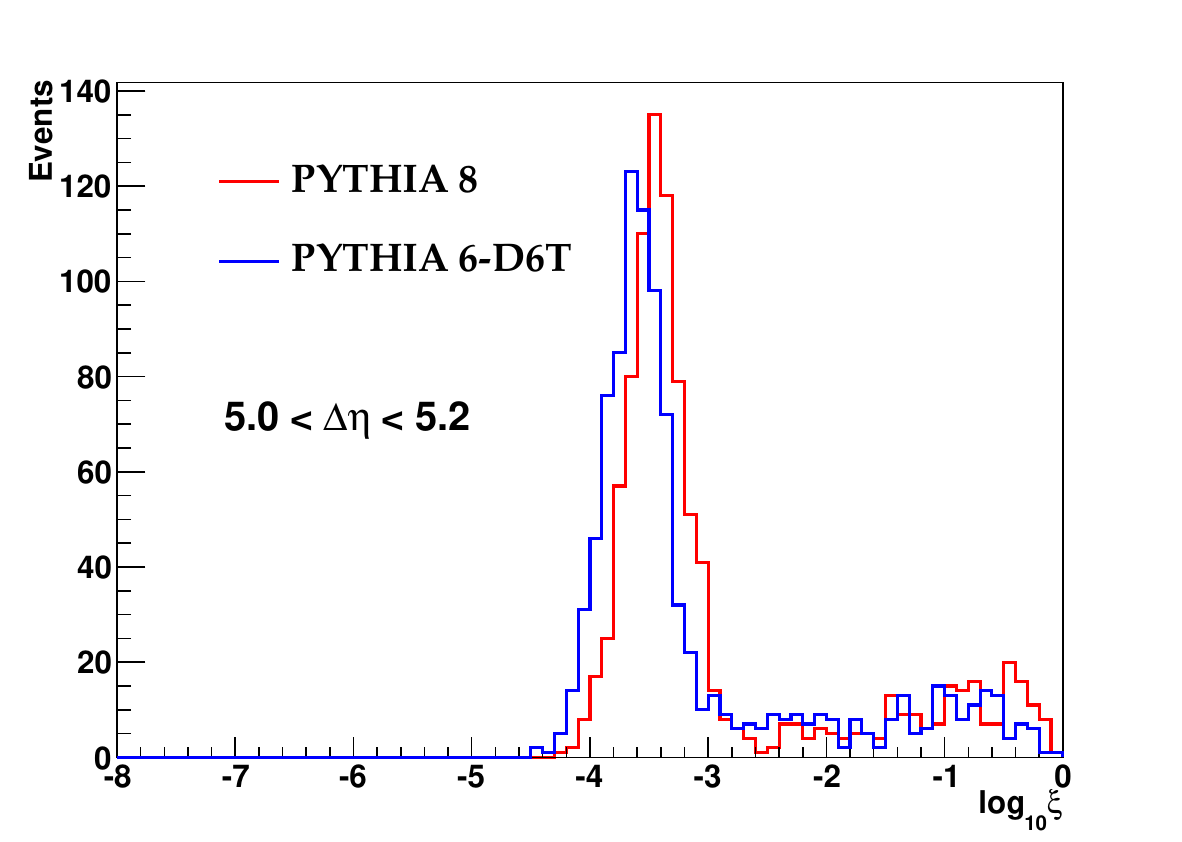}
    \caption{The range of $\xi$ values for SDD events with a gap size (a) \mbox{3.0 $<\Delta\eta<$ 3.2} and (b) 5.0 $<\Delta\eta<$ 5.2.
         Events are simulated by \mbox{PYTHIA 8} and PYTHIA 6-D6T, and only edge gaps are considered with a
         cut of $p_\text{T}>$ 200 MeV for the final state particles in $|\eta|<$ 5.2.}
\label{fig:Model}
  \end{center}
\end{figure}

As we discussed in the previous sections, a calculation of the diffractive mass can be made through its relation to the size of
the rapidity gap, however, this is model dependent. For a given gap size, the range of $\xi$ value can be different for different models.
As an example, we consider two different size of edge gap, 3.0 $<\Delta\eta<$ 3.2 and 5.0 $<\Delta\eta<$ 5.2, and calculate the diffractive mass of the
dissociated system for SDD events simulated by PYTHIA 8 and PYTHIA 6-D6T which use different set of parameters for 
the simulation~\cite{Sjostrand:2007gs, Sjostrand:2006za}.
The range of $\xi$ values and the difference in the range between models, are given in Fig.~\ref{fig:Model}.

\begin{figure*}[h!t]
  \begin{center}
    \includegraphics[width=0.42\textwidth]{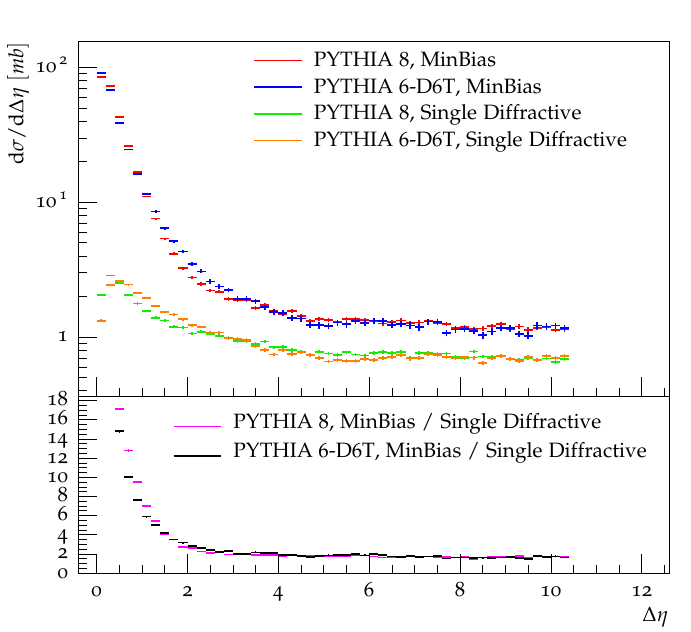}\hspace{0.5cm}
	\includegraphics[width=0.42\textwidth]{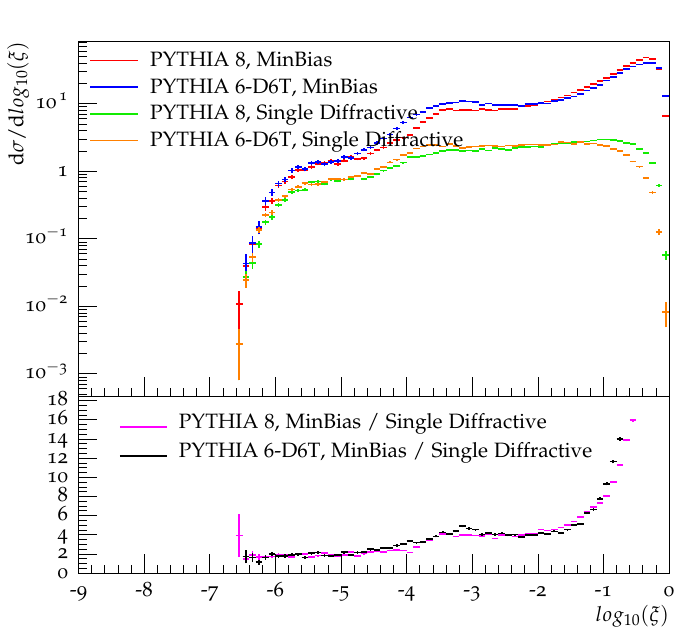}
    \caption{The distribution of $\frac{d\sigma}{d\Delta\eta}$ (left) and $\frac{d\sigma}{dlog_{10}\xi}$ (right)
         for different event classes simulated by PYTHIA 8 and PYTHIA 6-D6T.
         Only edge gaps are considered with a cut of $p_\text{T}>$ 200 MeV for all final state particles in $|\eta|<$ 5.2.}
\label{fig:Model2}
  \end{center}
\end{figure*}

We also study the model dependence of correcting an inclusive minimum bias measurement to one for SDD processes only
as presented in Fig.~\ref{fig:Model2}.
It looks like the difference is quite small especially in the $\Delta\eta>$ 3 region, but of course the difference could be larger
with other models.
The correction to get to SDD is large (about a factor of 2) and it is preferable to measure $\Delta\eta$ distribution
without performing such a correction, such that dependence on MC models is minimized.

\section{Conclusions}

Methods to select soft diffraction dissociation at the LHC experiments ATLAS and CMS are
studied by using large rapidity gaps in the events.
It is shown that the larger the rapidity covered, the more precisely measurements for diffractive dissociated events can be done.
In the limited detector rapidity coverage $|\eta|<5.2$,
one can select a sample of events of which 98.8\% are diffractive dissociated, according to PYTHIA 8, by requiring
an edge gap with a gap size $\Delta\eta>$ 3
and with a cut of $p_\text{T}>$ 200 MeV for all final state particles.
However, with this event selection, 42.1\% of the diffractive dissociated events will be double diffractive dissociated events and
it seems not possible to separate SDD from DDD events with a low diffractive mass by using edge gaps.
Central gaps look like a better candidate in order to try to distinguish SDD and DDD events, however, only a small fraction of
DDD events have a central gap.
Although the very forward detectors, ZDC, can provide a better distinction for the SDD and DDD events, it still seems not possible to select
a pure sample of SDD events within the detector limits.
Since it is not possible to make an unambiguous distinction between SDD and DDD events with the
limited rapidity coverage of the detector, $|\eta|<5.2$,
we should not attempt to make a measurement of SDD production.
It is preferable to measure the rapidity gap cross-section for all types of process, without attempting to correct for DDD events.
Also, due to the limited tracker range $|\eta|<2.5$, a primary vertex requirement in the event selection is not practical
to perform measurements for diffractive dissociated events. Particularly, very low mass soft diffractive events which
dissociate into the very forward rapidities and have a gap with a size of $\Delta\eta>$ 8, are suppressed by the primary vertex requirement.

\section*{Acknowledgements}

Valuable discussions with Hannes Jung are thankfully acknowledged.
One of us S. Sen would like to thank to University College London HEP group members for their kind hospitality.
This work was supported by the Marie Curie Research Training Network ``MCnet'' (contract number MRTN-CT-2006-035606), and 
by Hacettepe University, Scientific Research Projects Program through the Project: FBB-2015-6678.

\bibliographystyle{utphys}
\bibliography{svjourn3-epjc}

\end{document}